\documentclass{pasa}%

\title[The Zadko Telescope]{The Zadko Telescope: Exploring the transient Universe}
\author[Coward et al.]{D.M. Coward$^1$\thanks{david.coward@uwa.edu.au}, B. Gendre$^{2,3}$, P. Tanga$^4$, D. Turpin$^{5,6}$, J. Zadko$^1$, R. Dodson$^{7}$, M. Devog\'ele $^{8,9}$, E.J. Howell$^1$, J.A. Kennewell$^1$, M. Bo\"er$^3$, A. Klotz$^{5,6}$, D. Dornic$^{10}$, J.A. Moore$^{1}$ \and A. Heary$^{1}$\\
\\
\affil{$^1$School of Physics, University of Western Australia, M013, Crawley WA 6009, Australia}%
\affil{$^2$University of the Virgin Islands, John Brewers Bay, St Thomas, U.S. Virgin Islands 00802-9990, USA}%
\affil{$^3$CNRS - ARTEMIS, boulevard de l'Observatoire, CS 34229, 06304 Nice Cedex 04, France}%
\affil{$^4$Boulevard de l'Observatoire, CS 34229, 06304 Nice Cedex 04, France}%
\affil{$^5$Universit\'e de Toulouse; UPS-OMP; IRAP; Toulouse, France}
\affil{$^6$CNRS; IRAP; 14, avenue Edouard Belin, F-31400 Toulouse, France}
\affil{$^7$International Centre for Radio Astronomy Research, M468, The University of Western Australia,
Crawley, WA 6009, Australia}
\affil{$^8$Laboratoire Lagrange, Universit\'e C\^ote d'Azur, Observatoire de la C\^ote d'Azur, CNRS,
Blvd. de l'Observatoire, CS 34229, 06304 Nice cedex 4, France}
\affil{$^9$Institut d'Astrophysique, G\'eophysique et Oc\'eanographie, Universit\'e de Li\`ege, Belgium}
\affil{$^{10}$Aix Marseille Universit\'e, CNRS/IN2P3, CPPM UMR 7346, 13288, Marseille, France}}
\jid{PASA}
\doi{10.1017/pas.\the\year.xxx}
\jyear{\the\year}

\usepackage{graphicx}
\usepackage{epstopdf}
\usepackage{caption}
\usepackage{amsmath}
\usepackage{url}
\usepackage{natbib}
\usepackage{url}

\def \apj {ApJ}

\def \apjl {ApJL}
\def \mnras {MNRAS}
\def \aap {A\&A}
\def \aj {AJ}
\def \araa {ARAA}

\def \nat {Nature}

\def \pasa {PASA}

\def \planss {Plan. Sci.}

\newcommand{\degree}{$^{\circ}$}

\begin{document}%
\begin{abstract}
The Zadko Telescope is a 1 m f/4 Cassegrain telescope, situated in the state of Western Australia about 80 km north of Perth. The facility plays a niche role in Australian astronomy, as it is the only meter class facility in Australia dedicated to automated follow-up imaging of alerts or triggers received from different external instruments/detectors spanning the entire electromagnetic spectrum. Furthermore the location of the facility at a longitude not covered by other meter class facilities provides an important resource for time critical projects. This paper reviews the status of the Zadko facility and science projects since it began robotic operations in March 2010. We report on major upgrades to the infrastructure and equipment (2012-2014) that has resulted in significantly improved robotic operations. Secondly, we review the core science projects, which include automated rapid follow-up of gamma ray burst (GRB) optical afterglows, imaging of neutrino counterpart candidates from the ANTARES neutrino observatory, photometry of rare (Barbarian) asteroids, supernovae searches in nearby galaxies. Finally, we discuss participation in newly commencing international projects, including the optical follow-up of gravitational wave candidates from the U.S. and European gravitational wave observatory network, and present first tests for very low latency follow-up of fast radio bursts. In the context of these projects, we outline plans for a future upgrade that will optimise the facility for alert triggered imaging from the radio, optical, high-energy, neutrino and gravitational wave bands.
\end{abstract}
\maketitle%

\graphicspath{{figs/}{./}}
\section{INTRODUCTION}
\label{sec:intro}
\subsection{Transient source astronomy}
The Universe is teeming with fleeting transients, seen across the entire electromagnetic spectrum. In the optical, some last for only a few seconds, while others change slowly in brightness over years. Sources classed as transient include cataclysmic variable stars, blazars, stellar flares, hypernovae and supernovae, while many of the recently discovered sources are not easily explained. Some transients, such as gamma ray bursts (GRBs), have complex emissions spanning the gamma, x-ray and optical spectra, and are possibly strong gravitational wave and neutrino sources.

The 21st century has heralded a new trans-spectral era for exploration of the Universe: for the first time humankind is able to observe the Universe across the whole electromagnetic spectrum and beyond. Space based gamma ray burst detectors have opened a window into a Universe teeming with short-lived exotic phenomena, the so-called `transient Universe'. But to fully reveal the nature of transients requires near simultaneous observations from the radio, optical and high energy spectra, combined with data from the gravitational wave (GW) spectrum, initiated in late 2016 by the first observations of binary black hole mergers \citep{2016PhRvL_GW151226,PRL,O1_BBHs_2016}.

{\it Multi-messenger astronomy} spanning the entire electromagnetic spectrum has yet to be fully exploited because of the difficulty of rapid response automation, data processing, and co-ordination required between independent facilities. Despite the difficulties, progress has been made: we are entering a new dawn in the discovery and study of exotic transients. Space-based satellite detectors are being networked to optical telescopes for rapid follow-up imaging of transient targets. Furthermore, for the first time robotic telescopes in Australia, Chile and France are being networked together, performing automated imaging of transients analogous to a distributed parallel computing network, where imaging tasks are allocated to particular instruments. Another recent innovation is the coordination of optical and radio telescopes for coincident transient detection, which requires automated communication between different facilities to image the same sky (see section \ref{zfrb} for details).

\section{The Zadko Telescope}
The Zadko Telescope \citep{2010PASA...27..331C} is a 1 m f/4 Cassegrain telescope constructed by DFM Engineering Ltd, situated in the state of Western Australia  at longtitude, 115\degree 42' 49'' E, latitude, 31\degree 21' 24'' S and at an altitude of 50 metres above sea level. Situated about 70 km north of Perth \citep[see][for more detail]{2010PASA...27..331C}. It was donated to UWA by resource company Claire Energy, and opened by Australia's Chief Scientist on 1 April 2009, with a science goal of exploring a previously uncharted region of the `transient sky'. The most important imaging specifications of the Zadko telescope are outlined in Table \ref{tab_zadko}.

In 2010, the telescope control system was robotized using the the TAROT robotic telescope software system. This suite of independently running programs comprises two main components: AudeLA\footnote{\url{http://www.audela.org/english_audela.php}} and ROS (Robotic Observatory Software). There is also 3rd party software (mostly drivers) for the mechanical and sensors of the observatory (i.e., telescope + dome + dome interior + weather station). For the Zadko Telescope, The DFM supplied Telescope Control System is interfaced to ROS via the ASCOM ( Astronomy Common Object Model) protocol.

\begin{table}[tbh!]
\begin{center}
\caption{ Zadko Telescope imaging specifications }
\label{tab_zadko}
\begin{tabular}{ll}
\hline
CCD FOV & $23'.6\times23'.6$   \\
pixel scale & $2048\times2048$\,$13.5$\,$\mu$m\,square pixels\\
pixel size$^a$ &  $0.69$\,arcsec \\
filters & SDSS u, g, r, i  and clear\\
camera & Andor iKon-L  \\
CCD & DW436-BV back-illuminated sensor \\
mag limits & $m\approx21.5$ 180s exposure \\
phot accuracy & dispersion $< 0.01$ for $m=11$ \\
seeing & variable $(1-4)$ arcsec \\
\hline
\end{tabular}
\end{center}
\footnotesize{ \noindent
 $^a$ The CCD images employ $2\times2$ binning because of the seeing limitations, resulting in pixel sizes of about $1.4$\,arcsec}\\

\end{table}

A key feature of the telescope is the optimization for performing automated follow-up from alerts or triggers from different external instruments. In addition, it employs a dynamic scheduling system that allows other projects to operate when not in alert mode. Furthermore the location of the facility at a longitude not covered by other meter class facilities provides unique access to the Southern transient sky.

The aim of this paper is to review the status of the Zadko facility and science projects since it began robotic operation. Firstly we will report on several major upgrades to the infrastructure and equipment that have improved the capability of the facility for transient source follow-up. Secondly, we review the core science projects which include:
\begin{itemize} 

\item Gamma ray bursts: Rapid optical follow-up of {\it Swift} alerts

\item ANTARES neutrino alert follow-up

\item Solar system studies: Gaia satellite follow-up of asteroids and photometry of ``Barbarian asteroids''

\item Gravitational wave follow-up of alerts from aLIGO/Virgo

\item Nearby supernovae and lensed supernovae search


\item Emerging high priority project: automated low latency follow-up of radio transients (Fast Radio Bursts)

\item Education and training: space debris tracking (pilot project)

\end{itemize} 


\section{Zadko Telescope infrastructure and equipment upgrade}

The Zadko Telescope, which was first installed in June 2008, was initially housed within a 6.7 metre fibreglass dome. Due to the robotic nature of the telescope which can make hundreds of movements throughout the night, the demands placed on the fibreglass dome to continually adjust its position to accommodate these movements caused it to be in constant need of maintenance and repair and consequential failure.

In March 2011 it was decided to replace the fibreglass dome with a state of the art purpose built rolling roof facility with a dedicated constant temperature climate controlled operation room. In addition to this a climate controlled telescope Service room to mirror the Operation room was also added. This has the added benefit of allowing for possible future conversion into a second control room if further telescopes are added to the facility.

Due to the environmentally hostile and remote location of the telescope, it was decided to leave the telescope in its existing position during the decommissioning of the fibreglass dome and the construction phase of the new building. This was achieved by erecting a temporary watertight structure around the telescope before dis-assembly of the fibreglass dome. 

As the majority of the new structure is steel and prefabricated off-site the installation was performed in a relatively short time frame of approximately six months. The curved corrugated zinc-alumina roof was then installed without gutters so as to prevent leaf litter build up as a fire prevention strategy. The walls and ceiling panels of the Operation, Service and Telescope rooms are 100mm thick EPS core insulated panel which assist in maintaining reasonably constant temperatures throughout the year. Both Operation and Service rooms have split system air-conditioning units installed for constant temperature control.

\begin{figure}
\centering
\includegraphics[scale=0.5]{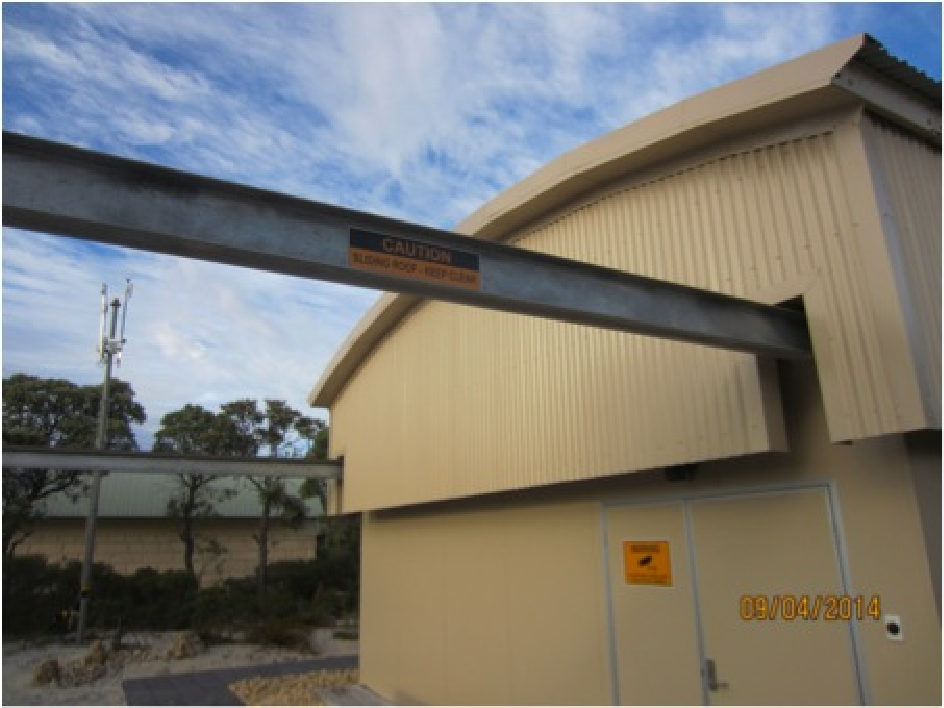}
\includegraphics[scale=0.38]{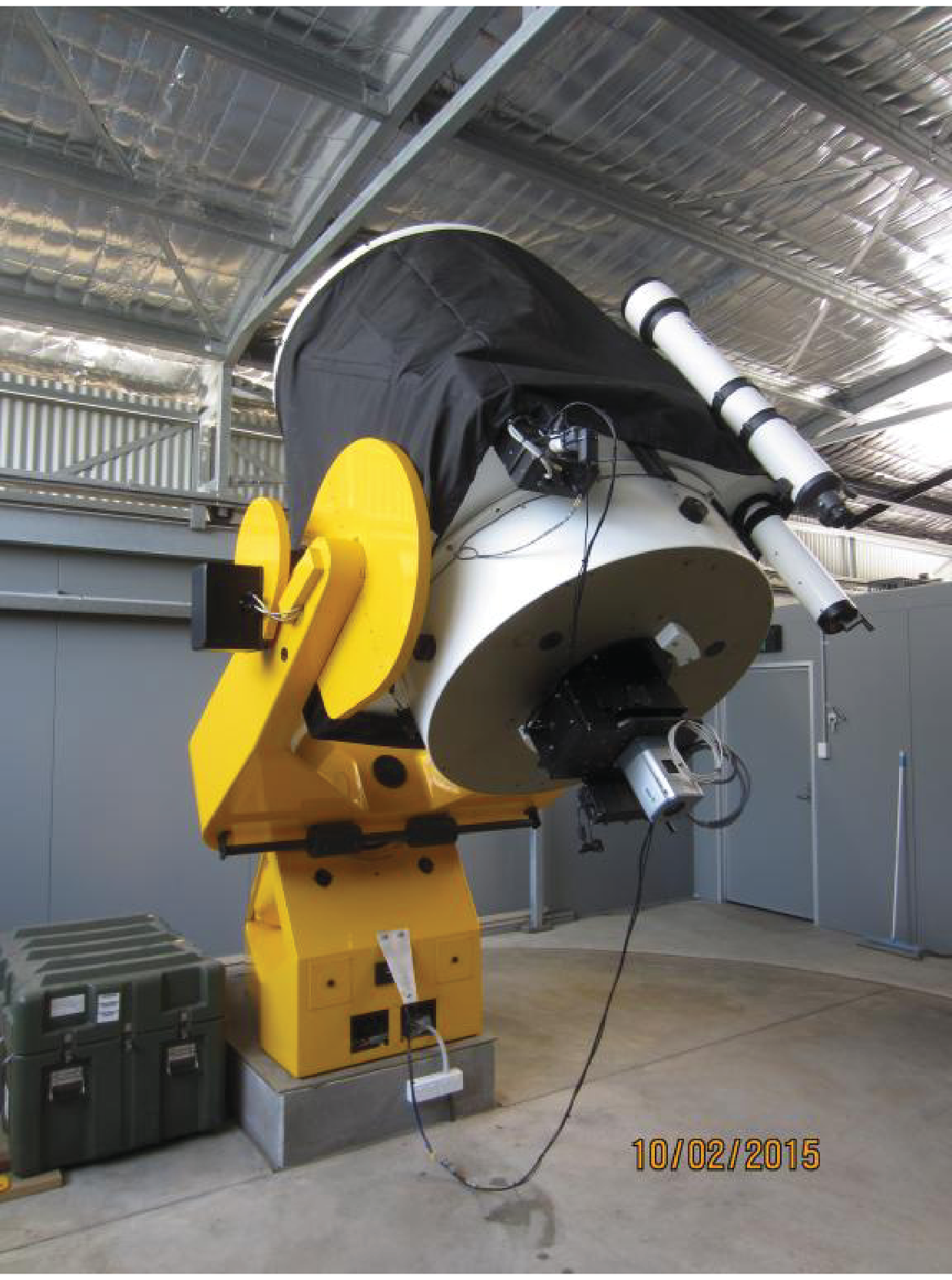}
\caption{{\bf Top} The new observatory constructed in 2012, with key features including fully automated slide off roof controlled by weather sensors interfaced to a PLC system. {\bf Bottom} The Zadko Telescope ready for operation. } \label{LF}
\end{figure}

The main control for the automatic operation of the rolling roof is provided by a PLC/Burgess system which was designed and installed by our French collaborators at Observatoire de Haute-Provence (OHP). This includes electrical and mechanical interfaces for temperature, humidity and wind sensors, flat field lamp, emergency stops and limit switches plus a main control panel.

An important feature of the building design was to protect the telescope from sudden adverse weather events. This was achieved by installing an external mast fitted with real time weather sensors (cloud, rain, humidity, wind, temp) which enables a signal to be sent to the PLC/Burgess. The rolling roof immediately responds to these signals and may close and reopen several times during the night according to the weather conditions. Approximate roof travel time to close/open is two minutes in either direction.

The main drive of the rolling roof consists of two three phase 0.75 kW brake flange motors connected to an in-line helical gear box with a 46:1 ratio installed on either side of the structure connected to a geared rack. The motors are powered by a variable speed drive (VSD) with single phase input and three phase output which is located within the PLC/Burgess cabinet. All critical drive components of the operation are supported by a Socomec 7kVA Uninterrupted Power Supply (UPS) in the event of an external power failure. These include the ability for the roof to close and for all computer systems to remain active (See Fig 1).


The existing Guide Acquired Module (GAM) which houses the Andor IKON-L camera has been redesigned to provide an improved field of view, provision for larger filters and to allow for future spectroscopy. This new GAM will be installed in 2016. Estimated downtime is six weeks to allow for both primary and secondary mirrors to be removed for re-coating and re-installation. A new flat field screen has been installed, suspended from the rolling roof at a distance and angle to achieve optimum results for camera calibration. Improvements to computer systems, data and internet provision are continuous and ongoing with major areas of concern being poor internet service and intermittent power outages.

\subsection{Environmental impact on equipment}
\begin{figure}
\centering
\includegraphics[scale=0.41]{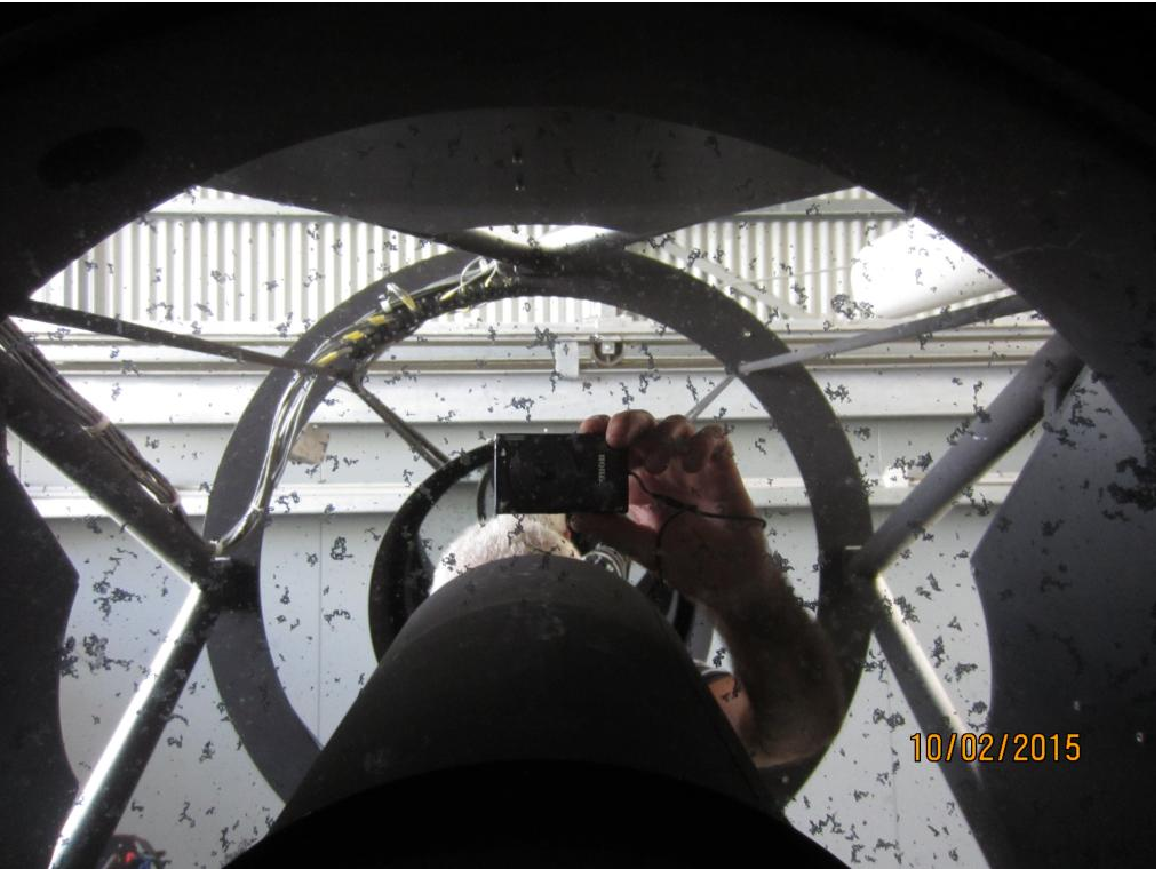}
\includegraphics[scale=0.41]{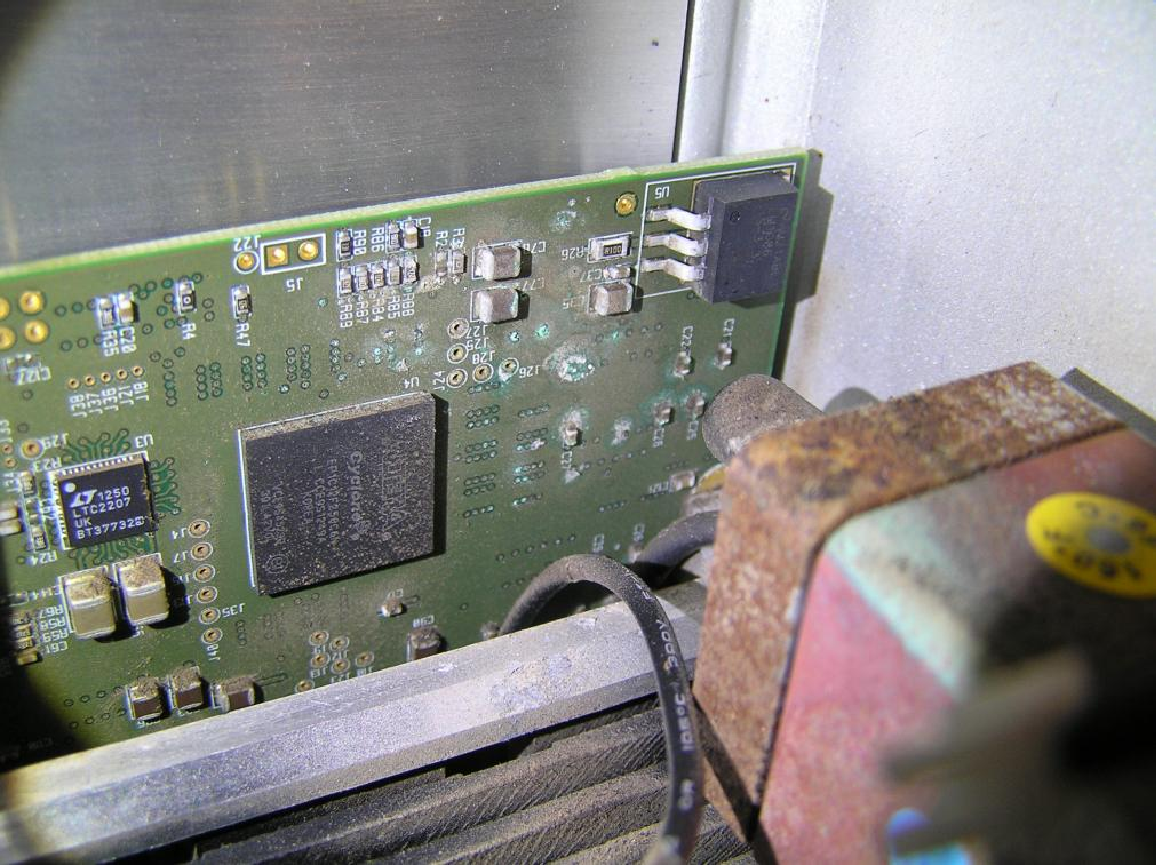}
\caption{{\bf Top} The primary mirror, showing corrosion of the coating (grey patterns) from the high levels of humidity and condensation, combined with dust and pollen.
{\bf Bottom} The main circuit board of the IKON-L CCD camera, which is exposed to the atmosphere, showing corrosion on all metallic components and connectors. This eventually led to a camera failure after about 14 months of use. } \label{mr}
\end{figure}

The observatory site has provided a very challenging environment for the equipment. In particular, humidity levels often rise to above 90\% during the night, mainly between June to September. Even though the system automatically shuts down at this humidity level, the primary mirror and camera (which is thermo-electrically cooled) are partially exposed. Condensation combined with dust, salt and pollen has accelerated the degradation of the mirror coating. To reduce this, the mirror has to be cleaned regularly on the time-scale of 4-6 weeks. Fig \ref{mr} shows the corrosion on the primary mirror coating and the affect of condensation on the main CCD camera circuit board. UWA technical staff are testing several filter options for the CCD that will inhibit dust intake and condensates on the circuit board.

\subsection {RAPIDO: Plans for a co-located wide field rapid response telescope}
The french leaders of the TAROT/Zadko projects have designed and constructed a prototype extremely rapid response telescope mount (slew speed 45 deg per second)\footnote{http://cador.obs-hp.fr/rapido/} . This mount is planned to control two off-the-shelf wide-field (combined 8-10 square degrees FoV) Takahashi (2 x 20cm) telescopes. The first imaging using a FLI Kaf16003 camera obtained a limiting magnitude of about 18 (clear filter). RAPIDO and Zadko will share the same robotic control system, which will enable RAPIDO to automatically send alerts to Zadko for deep follow-up of gravitational wave and fast radio burst candidates. The instrument is upgradable to accommodate a 0.4 m telescope, and will be co-located with the Zadko Telescope, subject to funding.

\section{Gamma-Ray Burst follow-up}

\subsection{Gamma-Ray Bursts}
\label{section_GRBs}

One class of extra-galactic transient that triggers the Zadko Telescope are Gamma-Ray Bursts (GRBs). They are the brightest electromagnetic explosions in the Universe \citep[see the review by][]{mes06}. Usually detected first in gamma and X-rays by NASA's {\em Swift} satellite \citep[][albeit other instruments can detect GRBs]{geh04}, they have distinct but poorly understood optical signals, including prompt emission, and a delayed afterglow.

The popular progenitor model for a GRB is either a massive stellar collapse, or a compact object merger triggering an explosion causing a burst of collimated gamma-rays powered by accretion onto the newly formed compact object. There are two classes of GRBs categorised by their durations and spectra: namely short-hard GRBs and long-soft GRBs. The former are supposed to be associated with the merging of compact binaries \citep{eic89} while the latter are expected to occur after the cataclysmic collapse of a massive star \citep{woo93}.

The popular description for GRB emission is the standard model, also termed the fireball model \citep{ree92, mes97, pan98}. In this framework, GRBs originate from the dissipation of kinetic energy of an ultra-relativistic outflow (with an initial Lorentz factor of 100-1000), decelerated by interaction with interstellar matter to produce a fading X-ray and optical afterglow. After the energy decreases to some threshold, the radiation beam is wider than the jetted out-flow, so the afterglow becomes observable from angles greater than the high-energy jet.

The fireball model has considerable deficiencies and challenges. For example, the energy budget deduced from the observed emissions and the measured distance implies an on-jet beamed geometry to keep the energy budget physically reasonable \citep{rho99}. However, {\it Swift} has put this geometry in question \citep{mes06}, allowing for exotic theoretical models \citep{geh08} that go-beyond standard relativity. Here, only coordinated multi-spectral observations starting a few seconds after the trigger can test theories.

\subsection{Zadko Telescope GRB afterglow observations}

The implementation of a robotic telescope for fully automated, calibrated imaging of GRBs is complex. The telescope must respond automatically to alerts sent by satellites through the GCN network \citep{bar94}, if all the environmental conditions are fulfilled. These conditions include the time of day (observations are at night), the position of the GRB on the sky, and the priority of the observation. Because of these limits, not all alerts can be followed-up.

Secondly, there is no manual intervention at the time of the automated triggered imaging to confirm if the sky is clear and the photometry is calibrated. About a third of the sources followed-up are not useful, either because of cloud cover, or because a technical problem corrupted the photometry. For the Zadko Telescope GRB afterglow follow-up campaign, we list the successfully imaged bursts in Table \ref{table_GRB}.

One can clearly see in the distribution of observation times `holes' corresponding to major refurbishments of the instrument, such as the installation of a new robotic observatory in 2012 (see above). Despite this, and other technical issues, the Zadko Telescope detected 11 GRBs in optical out of 40 successful follow-ups during its 6 year campaign. This is a typical rate for a robotic instrument \citep[see for instance][for a comparison with the TAROT instruments]{klo09}.

\begin{table*}
\caption{Zadko Telescope follow-up of GRBs. The localizations are obtained from the X-ray position, with an uncertainty of 1.5 arcsec, or the Swift Burst and Transient detector (denoted by a $^\gamma$), with an uncertainty of $1-4$ degrees. All times are given relative to the trigger time, and we report in this table only the upper limits (see Table \ref{table_mag} for the measurements). Zadko employs the same imaging strategy as TAROT: the first image is a 60 s exposure with tracking off (trailed images), followed by 30 s, 60 s 120 s and 180 s exposures in tracking mode. For late follow-up (hours post trigger), the 180 s exposures are stacked. The label 'ld' indicates that the data was lost during a computer update.} \label{table_GRB}
\centering
\begin{tabular}{ccccccc}
\hline
Burst & Triggering & Position             & Observation time & Detection & Upper &  Reference\\
      & satellite  & (RA-Dec)             & (Start - End s)  & (Yes/No)  & limit &       \\
\hline
081118 	& Swift    & 05 30 22 $-43$ 18 04          &   7056 - 10800  & Y & ---  & GCNC 8675 \\
090205 	& Swift    & 14 43 39 $-27$ 51 10          &  63360 - 64800  & Y & ---  & GCNC 8976 \\
090313	& Swift    & 13 13 36 $+08$ 05 50          &  15336 - 16920  & Y & ---  & GCNC 8996 \\
090509 	& Swift    & 16 05 41 $-28$ 23 06$^\gamma$ &  28800 - 29700  & N & 17.6 & GCNC 9363 \\
090516 	& Swift    & 09 13 03 $-11$ 51 16          &  16488 - 16524  & Y & ---  & GCNC 9380 \\
090927 	& Swift    & 22 55 54 $-70$ 58 49          &   7308 - 7848   & Y & ---  & GCNC 9956 \\
091127 	& Swift    & 02 26 20 $-18$ 57 08          &  66600 - 68040  & Y & ---  & GCNC 10238 \\
091221  & Swift    & 03 43 11 $+23$ 14 29          &  64195 - 161516 & N & ld & --- \\
100424  & Swift    & 13 57 49 $+01$ 32 19          & 153710 - 268123 & N & ld & --- \\
100628  & Swift    & 15 03 46 $+31$ 39 11$^\gamma$ &   7904 - 277914 & N & 21.0 & --- \\
100702  & Swift    & 16 22 46 $-56$ 32 56$^\gamma$ &  32801 - 302371 & N & ld & --- \\
100704  & Swift    & 08 54 34 $-24$ 12 10          &  23571 - 290112 & N & ld & --- \\
101011A & Swift    & 03 13 11 $-65$ 58 54          &   3924 - 4104   & N & 21.5 & GCNC 11336  \\
101024A & Swift    & 04 26 02 $-77$ 15 55          &  214.8 - 222    & Y & ---  & GCNC 11382 \\
120320A & Swift    & 16 10 04 $+08$ 41 47          &  24480 - 26640  & N & 20.7 & GCNC 13087 \\
120327A & Swift    & 16 27 28 $-29$ 24 54          & 149472 - 151200 & Y & ---  & GCNC 13164 \\
120422A & Swift    & 09 07 39 $+14$ 01 06          &  12960 - 17280  & Y & ---  & GCNC 13250 \\
121226A & Swift    & 11 14 34 $-30$ 24 23          &     63 - 335    & N & 19.4 & GCNC 14107 \\
130131A & Swift    & 11 24 31 $+48$ 04 34          &     71 - 5400   & N & 16.5 & GCNC 14161  \\
130206A & Swift    & 09 21 31 $-58$ 11 37          &    106 - 482  	 & N & 20.1 & GCNC 14185 \\
130313A & Swift    & 15 45 39 $-00$ 22 08          &     52 - 320    & N & 19.3 & GCNC 14295 \\
130315A & Swift    & 10 30 12 $-51$ 47 40          &   4392 - 5544   & N & 20.4 & GCNC 14330 \\
130408A & Swift    & 08 57 37 $-32$ 21 40          &  50364 - 52416  & N & 20.5 & GCNC 14372 \\
130427A & Swift    & 11 32 33 $+27$ 41 52          &  12420 - 13140  & Y & ---  & GCNC 14468 \\
130612A & Swift    & 17 19 11 $+16$ 43 11          &  33480 - 36720  & N & 21.2 & GCNC 14885 \\
130615A & Swift    & 18 19 19 $-68$ 09 40          &   5760 - 17280  & Y & ---  & GCNC 14906 \\
131205A & Swift    & 08 46 31 $-60$ 09 21          &  13680 - 16200  & N & 20.2 & GCNC 15579 \\
131218A & INTEGRAL & 07 35 07 $-64$ 44 14$^\gamma$ &  75168 - 76248  & N & 20.5 & GCNC 15603 \\
140129A & Swift    & 02 31 34 $-01$ 35 44          &  32472 - 293458 & N & 17.7 & --- \\
140301A & Swift    & 04 38 14 $-54$ 15 24          &     70 - 74880	 & Y & ---  & GCNC 16064 \\
140302A & Swift    & 16 55 26 $-12$ 52 42          &  34655 - 306638 & N & 17.7 & --- \\
140311A & Swift    & 13 57 13 $+00$ 38 31          &    120 - 1111 	 & Y & ---  & GCNC 15952 \\
140323A & Swift    & 23 47 50 $-79$ 54 16          &   3600 - 5760 	 & N & 20.8 & GCNC 16059 \\
140331A & Swift    & 08 59 28 $+02$ 43 02          &  29736 - 31248  & N & 20.1 & GCNC 16057 \\
140628A & Swift    & 02 42 40 $-00$ 23 06          &  24192 - 26172  & N & 19.9 & GCNC 16474 \\
140719B & Swift    & 02 38 55 $-02$ 33 02$^\gamma$ &     85 - 446 	 & N & 19.0 & GCNC 16610 \\
141017A & Swift    & 06 14 31 $-58$ 34 57          &   2520 - 2700	 & N & 22.4 & GCNC 16920 \\
141212A & Swift    & 02 36 30 $+18$ 08 49          &    123 - 712  	 & N & 20.3 & GCNC 17184 \\
150103A & Swift    & 08 46 40 $-48$ 53 09          &    122 - 347  	 & N & 18.9 & GCNC 17273 \\
\hline
\end{tabular}
\end{table*}

\subsection{Automated response time, detections and photometry}

\begin{figure*}
\centering
\includegraphics[width=16cm]{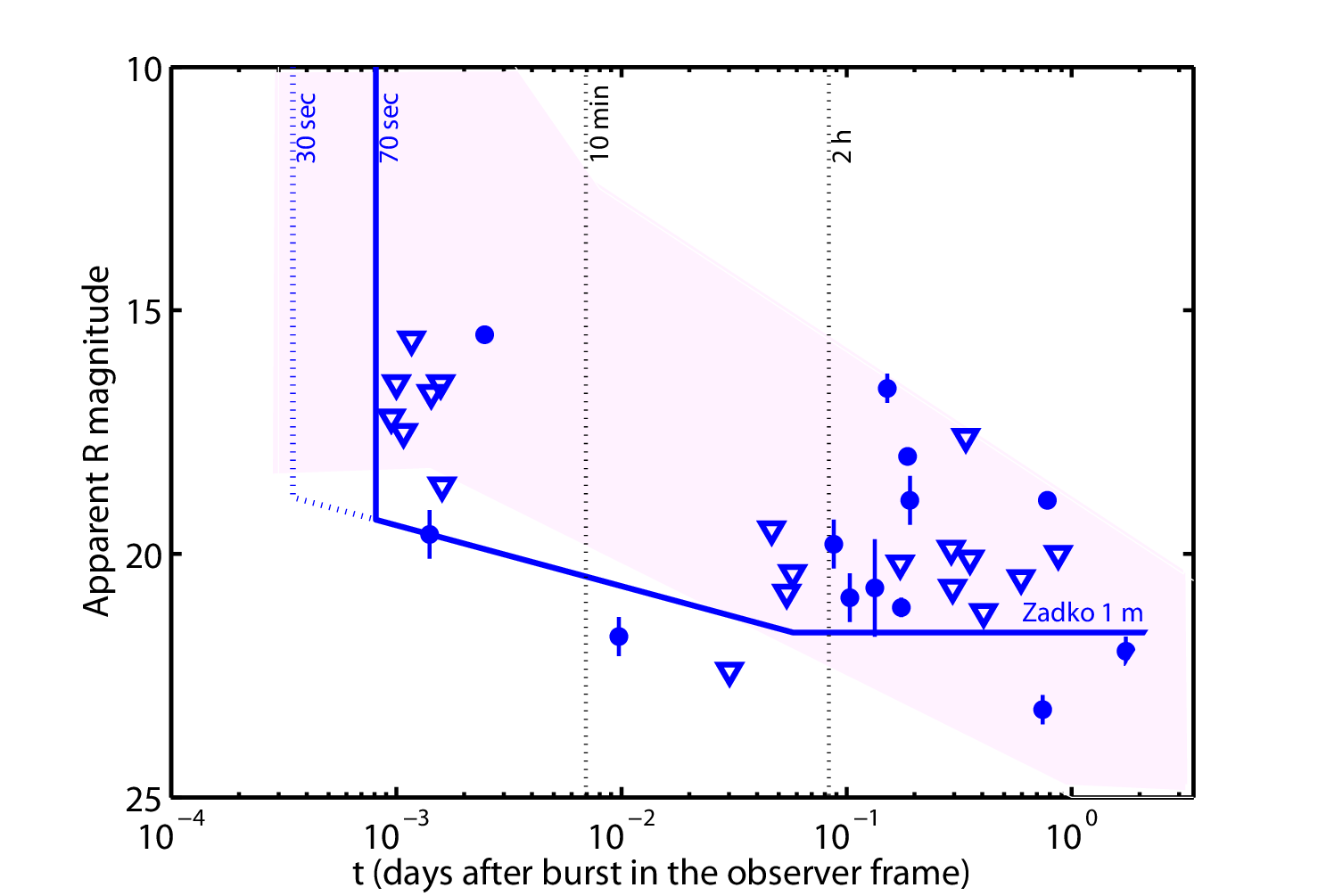}
\caption{Distribution of GRB afterglow observations and detections by the Zadko Telescope. GRB afterglows are represented by filled circles, while upper limits are represented as empty triangles. The red shaded area represents the light curves compiled in \citet{kan09}. The solid blue line is the theoretical observation limits under normal weather conditions with the previous observatory (terminated in 2011). The dashed blue line (at 30 sec) is the shortest response time possible for the new observatory (see the electronic version in colour).\label{fig_delay}}
\end{figure*}

The response time is a key parameter for all robotic telescopes. It will constrain the main scientific drivers in a given field. Comparing the shortest possible response time with the distribution of GRB durations \citep{kou93}, it is clear that the Zadko Telescope cannot be used to study the early prompt phase of normal long GRBs, with a mean duration of the order of 20 seconds, but can image the dynamic early afterglow phase during the transition from prompt to afterglow. We highlight that the response time has improved to a minimum possible response time of $>30$ s, during the last 2 years because of the new observatory (see above).

The 90\% confidence level limiting magnitude reached by the Zadko Telescope is consistent with theoretical expectations shown as the thick blue line\footnote{This limit is computed for normal weather conditions, but exceptionally good weather can lead to deeper (more sensitive) observations, as shown in the Figure.} in Fig. \ref{fig_delay}. It clearly indicates that the telescope cannot probe the entire GRB afterglow flux/luminosity distribution. One day after the burst, only half of the afterglows can be detected. After three to four days, Zadko cannot detect any afterglows. As a consequence, this constrains the observation strategy against long follow-up delay times (one day or more) of the afterglow, and highlights that the optimal follow-up strategy is a low latency response (comparable to the reaction time). At one day post alert, it is more practical to search for the host galaxy using deep (stacked) exposures.

In Fig. \ref{fig_delay}, the point representing GRB 090205 is an outlier located below the detection limit. However, it was observed during the commissioning phase with a different camera and settings than the one used after robotization: the detection limit presented in the figure does not apply to this point.

The Zadko Telescope can be used to study so called ``dark'' GRBs \citep{jak04}. This subset of GRB afterglows reveal a lack of emission in optical when compared to the X-ray, and their nature still remains puzzling. Because Zadko can obtain upper limits during the first minutes of the burst (up to about 20, see Fig. \ref{fig_delay}), the data can be used to classify a burst as dark (the faintness observed in optical can also be linked to a global faintness observed at all wavelengths), and provide a secondary trigger for a deep follow-up of the GRB location and host galaxy.

Among the sample, several afterglows were imaged, but (for most of them) only for a short duration (typically for about one to two hours). In addition, most sources were visible with a delay of several hours. Only four events were followed-up a few tens of seconds after the trigger and lasting for more than 3 hours: GRB 090313, GRB 101024A, GRB 130427A, and GRB 140311A. We list all the observations and detections in Tables \ref{table_GRB} and \ref{table_mag} respectively. It is beyond the scope of this paper to present a detailed analysis of each burst observed by the Zadko telescope, but more detailed studies have been undertaken \citep[e.g.][for GRB 101024A]{gen11}.

\begin{table}
\caption{GRB afterglows imaged by the Zadko Telescope. Filters are labeled 'C' for clear and 'R' for R band. For clear filter images, the magnitude is expressed in equivalent R band for comparison with other instruments. The time given relative to the trigger time, and the imaging strategy is the same as described in Table \ref{table_GRB}. \label{table_mag}}
\centering
\begin{tabular}{cccc}
\hline
Burst & Time    & Filter & Magnitude\\
      &  (s)    &        &      \\
\hline
081118 	&  8856 &   R    &  $20.9 \pm 0.5$ \\
\hline
090205 	& 63360 &   C    &  $23.2 \pm 0.3$ \\
\hline
090313	& 21346 &   C    &  $18.02 \pm 0.08$\\
090313	& 22662 &   C    &  $18.5 \pm 0.1$  \\
090313	& 24115 &   C    &  $18.5 \pm 0.1$  \\
090313	& 26028 &   C    &  $18.62 \pm 0.09$\\
090313	& 27890 &   C    &  $18.78 \pm 0.09$\\
090313	& 29854 &   C    &  $18.8 \pm 0.2$  \\
\hline
090516 	& 16500 &   C    &  $18.9 \pm 0.5$ \\
090516 	& 17101 &   C    &  $19.5 \pm 0.3$ \\
090516 	& 17760 &   C    &  $20.3 \pm 0.2$ \\
\hline
090927 	&  7560 &   R    &  $19.8 \pm 0.5$ \\
090927 	&  9480 &   R    &  $20.0 \pm 0.5$ \\
\hline
091127 	& 67440 &   R    &  $18.9 \pm 0.1$ \\
\hline
101024A & 218.2 &   C    & 16.6 $\pm$ 0.3  \\
101024A & 224.6 &   C    & 16.7 $\pm$ 0.3  \\
101024A & 232.1 &   C    & 16.9 $\pm$ 0.3  \\
101024A & 240.7 &   C    & 17.3 $\pm$ 0.3  \\
101024A & 274   &   C    & 17.6 $\pm$ 0.3  \\
101024A & 320   &   C    & 17.8 $\pm$ 0.3  \\
101024A & 364   &   C    & 18.0 $\pm$ 0.3  \\
101024A & 409   &   C    & 18.0 $\pm$ 0.3  \\
\hline
120327A &150336 &   C    &  $22.0 \pm 0.3$ \\
\hline
120422A & 15120 &   C    &  $21.1 \pm 0.2$ \\
\hline
130427A & 12780 &   R    &  $15.5 \pm 0.1$ \\
130427A & 13560 &   R    &  $15.5 \pm 0.1$ \\
130427A & 14640 &   R    &  $15.4 \pm 0.1$ \\
130427A & 15480 &   R    &  $15.6 \pm 0.1$ \\
130427A & 16260 &   R    &  $15.5 \pm 0.1$ \\
130427A & 18690 &   R    &  $15.6 \pm 0.1$ \\
130427A & 26910 &   R    &  $16.3 \pm 0.2$ \\
\hline
130615A & 11520 &   R    &  $20.7 \pm 1.0$ \\
\hline
140301A &   841 &   C    &  $21.7 \pm 0.4$ \\
\hline
140311A &   223 &   R    &  $19.6 \pm 0.5$ \\
140311A &   282 &   R    &  $19.9 \pm 0.3$ \\
140311A &   427 &   R    &  $19.2 \pm 0.3$ \\
140311A &   821 &   R    &  $18.5 \pm 0.3$ \\
\hline
\end{tabular}
\end{table}

\subsection{Further GRB science opportunities}

The response time and limiting magnitude set the potential for breakthrough science in relation to testing the standard fireball model. Potential new science and discoveries from Zadko Telescope automated triggered imaging include:

\begin{itemize}
\item The spectral evolution of the prompt phase. The Zadko telescope will not compete with faster robotic instruments for observing the start of the prompt phase, as can be noted in Fig. \ref{fig_delay}, however, its response time and sensitivity could probe the origin of the early optical emissions prompt phase of the brightest and longest duration GRBs.

\item The plateau phase. One of the main results from the {\em Swift} satellite is the presence of a plateau phase seen in X-ray between the prompt and the afterglow phases. The nature of this plateau is debated, and the comparison with optical data is critical to test the current models. \citet{boe15} has shown that the distribution of the end of the prompt phase in X-ray peaks at about 100 seconds. Accordingly, \citet{wil07} have shown that this plateau phase ends about $10^4$ seconds after the trigger. Thus, there is a significant fraction of bursts with the plateau phase starting within the window of Zadko telescope triggered imaging times.

\item Population III stars. The first stars, are metal-poor and proposed relics of the early Universe. They should form at very large redshift \citep[z $\sim 10-20$,][]{bro06}, but could rarely be present at $z \sim 2-4$ \citep{tom07}, and should produce GRBs \citep{suw11, nak12}. Moreover, isolated blobs of original matter, without metals, could be present near the Milky Way, and could form, if disturbed, new population III stars \citep{mor11}. The properties of these GRBs are expected to be similar to ultra-long GRBs \citep{gen13, mac13}, with a prompt phase lasting more than 600 seconds (indicated by the 10 minutes line on Fig. \ref{fig_delay}).
\end{itemize}

\section{Solar system science -- asteroid photometry}
\label{sec:SSO}

Measuring brightness variations is the primary source of information on an asteroid's shape. Photometric light curves also provides other very useful  physical parameters, such as the rotation period and the pole coordinates. Determination of these physical properties requires photometry at different geometries, i.e. over several oppositions \citep{2015MNRAS.453.2232D}.

Combining these observations with other measurements of the size (for instance by thermal modelling or stellar occultation), a more accurate volume, and density, can be determined \citep{2012P&SS...73...98C}. Also, shape is an indicator for the past collisional history of an asteroid. Without knowledge of an asteroid's shape, application of thermo-physical models to derive thermal inertia and size, is affected by significant uncertainties.

Currently, only the brightest asteroids have been systemically observed during several oppositions.
This is mainly the result of the use of a large number of small telescopes i.e. $20$ to $40$ cm \citep{2014ExA....38...91M}. Photometry is also obtained as a by-product of other observations, such as asteroid astrometry, or large sky surveys. In such cases, however, complete light curves are generally not obtained, but only sparse photometry. While it has been shown that sparse data can be exploited for the goals mentioned above, the quality of the data requires a careful selection and the process does not work for all the objects (Hanus et al. 2013). As of late 2015, $\sim$400 detailed asteroid shapes\footnote{\mbox{see e.g. the DAMIT data base:}\url{http://astro.troja.mff.cuni.cz/projects/asteroids3D/web.php}} \citep[see][using $\sim$7$\times$10$^5$ known asteroids]{durech2010} and $\sim$2000 rotational period measured, with rough shape parameters \citep{harris2012}.

Unfortunately, very few medium size telescopes (meter-class) are systematically exploited for asteroid photometry. In comparison to the largest telescopes (4m and more) they can offer more flexibility of use and are generally available for a longer time. Also, their sensitivity considerably extends the number of asteroids available for photometry. The Zadko telescope is ideal for performing sensitive photometry using the robotic scheduler.

Another important issue for asteroid observations is the measurement of accurate position in the sky. Several asteroids are usually in the field of view of the Zadko Telescope, when pointing near the ecliptic position. Reporting of the astrometric measurements of such asteroids will lead to an improvement of the asteroid orbit database. Furthermore, Zadko is part of a network of facilities that will follow-up Gaia (ESA satellite for astrometry and photometry) asteroid alerts in the coming years \citep{2013PASA...30...14T}. In parallel with this work, new models that constrain optimal search areas, and imaging cadences for narrow and wide-field survey telescopes was investigated \citep{2012MNRAS.420L..28T,2012MNRAS.424..372T}.
We detail below recent programs in active development employing the Zadko telescope for asteroid photometry.

\subsection{L-type and Barbarian asteroids}

Asteroids belonging to the L/K taxonomic classes could hold the key to understand the early phases of formation of the Solar System, when the solid phase in the proto-planetary nebula, which was largely dominated by highly refractory material, which solidify at the highest temperatures. Recent numerical simulations \citep{johansen2011} suggest that, in those conditions, typical of the
first $\sim$1 Myr after the disk formation, local concentrations of solid grains could have been rapidly assembled in large planetesimals of $\sim$100 km. Recently, near-IR observations \citep{Sunshine08} have suggested that some asteroids could also include in their composition a high fraction of that primordial material, hence be extremely ancient. A large network led by P. Tanga recently started an extensive activity of physical characterization of those asteroids, the so-called Barbarians, from the name of (234) Barbara, the first of this type discovered \citep{cellino_strange_2006}. The anomaly leading to the identification of the Barbarians was the peculiar variation of their linear polarization with the phase angle (the angle between the directions to the Sun and to the Observer, as seen from the asteroid).

13 Barbarians have been found so far \citep{gil-hutton_new_2008, masiero_polarization_2009, Cellinoetal2014}. Seven of them are in the Watsonia dynamical family, caused from the disruption of a single parent body \citep{Cellinoetal2014}. In terms of taxonomy, based on their similar spectro-photometric properties, Barbarians are classified as members of the unusual classes: $L$, $Ld$, and $K$.

 In absence of solid theoretical models capable of explaining the observed polarization, \citet{cellino_strange_2006} suggested, as a speculative explanation, that anomalous polarization could also be due to large-scale concavities responsible for causing an anomalous distribution of incident sunlight and scattering angles. In this respect, (234) Barbara has a very irregular shape and reveals large concavities, as shown by \citet{delbo_VLTI_2009}, and later confirmed by a more extensive analysis using different techniques \citep{tangaetal2015}.

For the small sample of known rotation periods of the largest Barbarians (7 objects), all of them are slow rotators (period 12-24h or more) with respect to the average period of similar-sized asteroids (5-6h). Slow rotation periods and the possible presence of concavities are indicators of past collisional history. However, the verification of this hypothesis requires a large amount of data on rotations and shapes. One tool for determining rotation periods and indications of shape irregularities is to obtain light curves spanning a near complete period.

\subsubsection{Zadko photometry of asteroids}
As the rotation period of the so called {\it slow rotators} generally exceeds the duration of night on Earth, telescopes distributed across many longitudes are required.  Ideally a light curve should be obtained over $\sim$1 week, otherwise the changing illumination/observation geometry will make its interpretation more complex.
For these reasons, Zadko is exceptionally well placed relative to other instruments that have also been employed (in France, Poland, USA, Chile). We note that this type of observation requires blocks of dedicated time, several hours of imaging spread over a night. To achieve this task automatically on the Zadko Telescope, we created a high priority schedule that could only be interrupted by a GRB alert.

\begin{figure*}
\centering
\includegraphics[width=12cm]{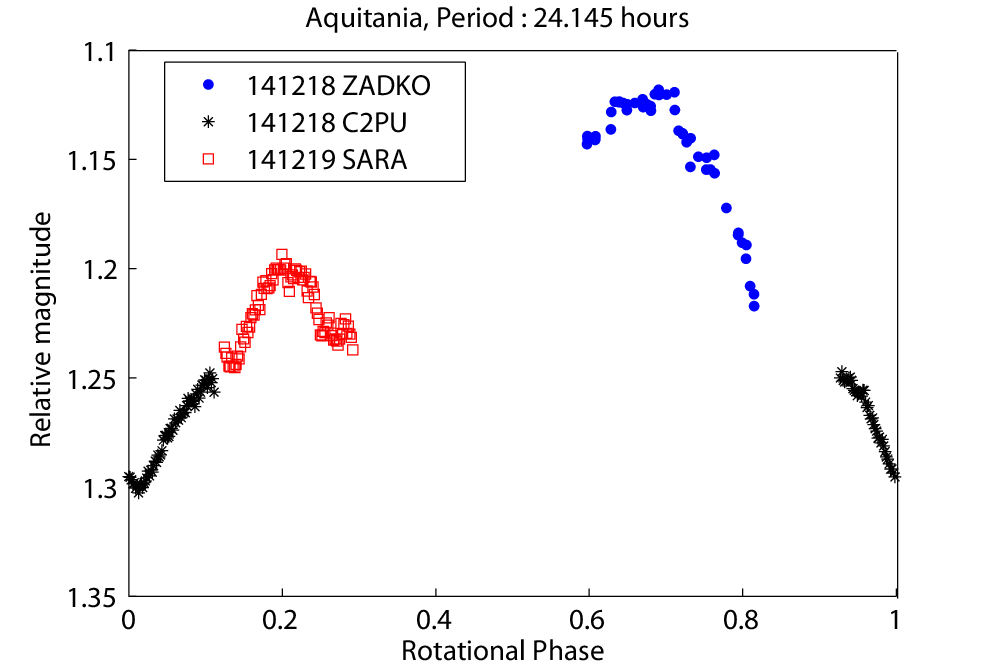}
\caption{Lightcurve of (387)~Aquitania. The observations by the Zadko Telescope provided light curve measurements around the brightness peak, unobserved at other longitudes. With a rotation period close to 24h, a single site can only acquire partial light curves.}
\label{Aquitania}
\end{figure*}

During one year (2014 to 2015), the Zadko telescope contributed to the period determination for 11 asteroids. These objects, (and most of the known asteroids) do not have measured rotation periods, or at best preliminary ones with unconfirmed values (with accuracies of the order of 0.1 days or worse) obtained from partial light curves not covering full rotations. Fig.~\ref{Aquitania} illustrates a typical example, in which Zadko observations covers a portion of a light curve observable only from Australia, thus capturing the maximum of brightness during a single rotation. The European longitudes of C2PU close to Nice, France ensured the coverage of the minimum over the same rotation of the asteroid.

By obtaining the near complete light curves, we were able to measure rotation periods with uncertainties better than 1$\times$10$^{-3}$-1$\times$10$^{-4}$ days with a measured dispersion $<$0.01 magnitudes for targets of relatively high brightness (V$\sim$11-13). This accuracy exceeds the photometric accuracy required for deriving most asteroid rotation peroids (having typical amplitude variation of 0.2-1 mag). Zadko photometry contributed to the shape determination of (234) Barbara \citep{tangaetal2015}, and updated results are due to appear in a paper in preparation.

\section{ANTARES follow-up}
\label{sec:Ant}
ANTARES is the largest high-energy neutrino telescope in the Northern Hemisphere, \citep{ageron2011}, and is aimed at the search for a muon neutrino signal from astrophysical sources. Such a discovery would unambiguously probe the identified source as an efficient cosmic accelerator of hadrons which is a key to solve the mystery of the ultra high-energy cosmic-ray (UHECR) origin in the universe. A multi wavelength follow-up of the neutrino alerts has been developed by the ANTARES collaboration \citep{ageron2012}, in order to quickly detect the electromagnetic counterpart from a potential neutrino source from radio to gamma-rays. This program denoted as TAToO (Telescopes-ANTARES Target of Opportunity) provides, in real-time using a socket protocol similar to the GCN system, the coordinates of the most promising neutrino events detected by the ANTARES telescope.

\begin{figure*}
\centering
\includegraphics[trim = 0 0 80 500,clip=true,width=0.8\textwidth]{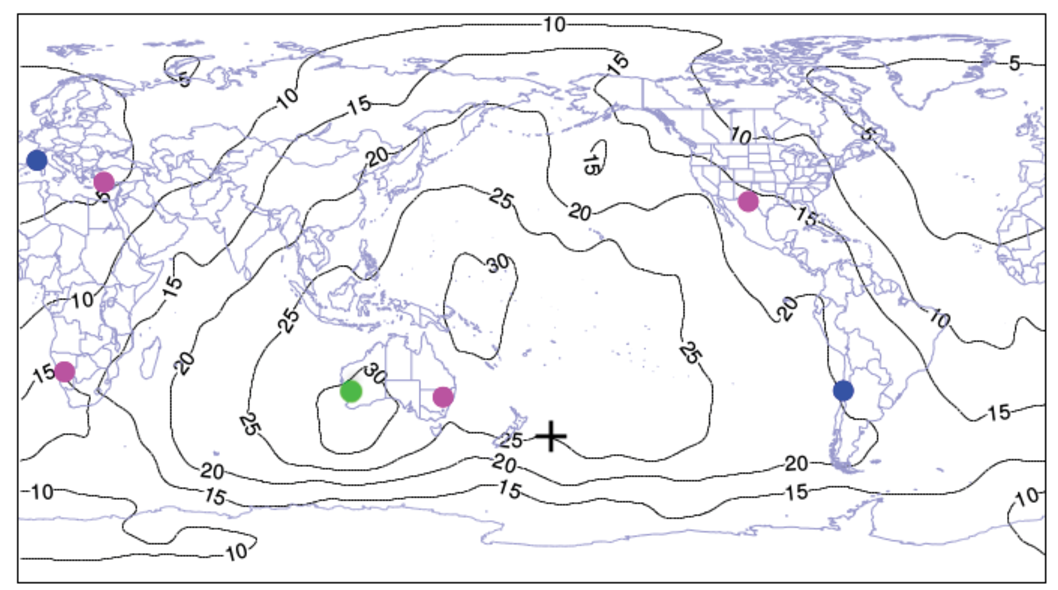}
\caption{The contours on the world map indicate the percentage of neutrino triggers visible at low latency.
(based on 140 ANTARES alerts). The black cross indicates the antipodal point for the ANTARES experiment. The TAToO robotic follow-up network is shown during the period 2012-2014 : (blue) ROTSE telescopes, (magenta) TAROT and Zadko telescopes, (green).}
\label{fig:map}
\end{figure*}

In the optical domain the Zadko Telescope is well matched to the TAToO program for several reasons :
\begin{enumerate}
\item It is located where the maximum fraction of ANTARES alerts are visible, see Fig.\ref{fig:map}.
\item Because of its rapid response to the alert system it can quickly observe the neutrino alert position in the sky. This feature is very important if the neutrino signal is produced by a rapidly fading transient source such as a GRB.
\item The limiting magnitude of the Zadko telescope ($\sim 21$ in 60s of exposure in a clear night without a bright moon) can place stringent limits on the optical brightness of the potential neutrino source.
\end{enumerate}

From 2013, the Zadko Telescope has participated in the robotic telescope network of the TAToO program, and succeeded in responding to 58 neutrino alerts. Among these, 3 were rapidly followed-up in less than 160 seconds after the neutrino trigger (ANT140301A : 64s, ANT140323A : 155s and ANT141220A : 53s) see Table. \ref{tab:stat_delay_ANT}. Unfortunately, no new transient sources were observed, but upper limits in R magnitude were derived for ANT140301A (C$\ge$19.8) and ANT141220A (C$\ge$20). No reliable limiting magnitude could be obtained for ANT140323A images because of their poor quality. In the framework of GRBs, the Zadko Telescope is able to exclude a typical bright GRB origin for the 2 neutrinos ANT140301A and ANT141220A, because most GRB afterglows are typically brighter at 60s than the limits we measure, see Table \ref{tab:stat_delay_ANT}. However, the GRB origin cannot be completely excluded because the GRB progenitor could also be intrinsically sub-luminous, or embedded in a dense circum-stellar medium. For the 55 late follow-up alerts ($\ge$ 1day after the neutrino trigger), no new transient source was discovered and no strong constraints on the GRB origin could be set because the major part of the GRB afterglow population is below the limiting magnitude of the Zadko Telescope 1 day after the $\gamma$-ray prompt emission.

As shown in Table.\ref{tab:stat_delay_ANT}, the response time of the Zadko Telescope can exceed 2 days. This long response time is due to observational constraints (the field of view is not immediately observable due to its position in the sky, the field of view is too close to the moon) or bad weather conditions (clouds, rain, wind). However, a problem in the scheduler program was detected and prevents us from following-up 7/55 alerts with shorter reaction time. Because these alerts were received during the day, the first observation was reported to the next night, i.e one day after the receipt of the ANTARES alert. Thus the reaction time of the Zadko telescope was biased and no alert could be followed-up in a delay of few minutes to few hours. This problem was solved in February 2015 with an update of the observational strategy of the ANTARES alert follow-up. The new program is now fully optimised (in response time) to detect a potential optical counterpart.

\begin{table}
\caption{Zadko follow-up of the ANTARES alert in the period 2014-2015. The reaction time corresponds to the time between the moment when the neutrino is detected by the ANTARES detector and the time when the first image is taken by the Zadko telescope. In the case of short time delays the limiting magnitude of the image was calculated.\\
\\
($^\bigstar$) Reaction time is biased ($\sim$ 1 day offset) because of an error in the scheduler program
\label{tab:stat_delay_ANT}
}
\centering
\begin{tabular}{ccc}
\hline
alert & Reaction time  & New detection\\
      &	   	    &	Yes/No \\	
\hline
ANT140123A 	&  2.12808 day &   N \\
ANT140125A$^\bigstar$ 	& 1.36891 day &   N \\
ANT140126A$^\bigstar$	& 0.96329 day &   N\\
ANT140202A$^\bigstar$	& 1.25782 day &   N  \\
ANT140227A	&  6.65249 day  &   N  \\
ANT140228A 	& 5.80076 day &   N\\
{\bf ANT140301A}	& {\bf 64.412 sec} &   N (R$\ge$19.8)\\
ANT140304A$^\bigstar$	& 1.32959 day &     N  \\
ANT140309A$^\bigstar$ 	&  1.45869 day &   N \\
ANT140311A 	& 2.30843 day &   N \\
{\bf ANT140323A}  	& {\bf 154.976 sec} &   N (--) \\
ANT140505A 	&  59.96 day &   N \\
ANT140619A$^\bigstar$ 	&  1.11634 day &   N \\
ANT140630A 	&  2.33353 day  &   N \\
ANT140818A  & 45.3101 day &   N \\
ANT140914A &  5.29319 day &   N \\
ANT140925A & 5.20087 day &   N \\
ANT141027A & 6.16642 day &   N  \\
ANT141112A  & 6.82595 day   &   N  \\
{\bf ANT141220A} & {\bf 52.896 sec}   &   N (R$\ge$20.0)  \\
ANT150111A &  4.31658 day   &    N  \\
ANT150122A & 6.49385 day   &    N  \\
ANT150131A &  9.23907 day &   N \\
ANT150213A & 4.26905 day  &   N \\
ANT150224A$^\bigstar$ & 1.43219 day &   N \\
ANT150321A & 15.7736 day &   N \\
ANT150324A & 10.5152 day &   N \\
\hline
\end{tabular}
\end{table}

\section{Supernovae search}

Supernovae (SNe) are energetic transients related to either binary systems, implying an accreting white dwarf \citep[Type Ia, see][for a recent review]{whelan1973,nomoto1982,maoz2014} or the collapse of massive stars \citep[Type II-Ib-Ic, see][]{janka2012}. Type Ia SNe are widely used as cosmological probes \citep{perlmutter1999,riess1998} as they can represent standard candles. Core collapse supernovae (CCSNe) still raise many unanswered questions concerning their origin, including: what are the properties of the stellar progenitors, and those of the newly born compact stellar remnants? What is the efficiency of the nucleosynthesis in massive stellar explosions? What would be the gravitational wave signal from a stellar collapse? Since the discovery of the first neutrino signal from the core collapse SN1987A, and more recently the discovery of the X-ray/UV signature of the shock breakout in SN2008D, \citep{soderberg2008}, type II supernovae are an active research area.

Because SNe are relatively rare events in the Universe (about 1 SN/gal/100yr), the typical strategy to detect them consists of observing a large number of galaxies each night which implies scanning a large part of the sky each night. Because of this strategy, and the growing number of facilities, about 200-300 SNe have been discovered each year for the last 15 years. However, because these discoveries often occur when the supernova is already in its declining phase, the features of the early rising supernovae are missing.

 As shown in \citet{soderberg2008}, in the case of CCSNe, detailed observations of the early phase of the supernovae (few hours to few days after the explosion) will help to understand the explosion mechanism, and to put constraints on the photospheric radius of the stellar progenitor, which is one of the key parameters to understand the phenomenon. It would also help to improve numerical simulations of CCSNe, as there is sparse data covering this early phase. Finally, observing the early stage of the SN explosion is important to estimate the total luminosity, the age of the SN, and the early time evolution of the SN spectrum.

\subsection{Zadko supernova follow-up}
Because the field of view of the Zadko Telescope cannot scan a large part of the sky, an innovative SN search strategy has been developed. The goal is to detect SNe during the early stage of the explosion, and to follow-up until they reach their maximum brightness ($\sim$ 20 days after the explosion) with a high sampling. For some fraction of CCSNe, we expect to detect the tail of the thermal shock breakout emission in the optical/near-UV domain ($\sim$ 1 day after the explosion) as already observed for a few of CCSNe, SN1987A, \citep{hamuy1990}, SN1993J, \citep{richmond1994}, SN1998bw, \citep{galama1998}, SN1999ex, \citep{stritzinger2002} and SN2008D, \citep{modjaz2009}. To achieve this, we designed a catalog of nearby galaxies of interest which historically are prolific in producing SNe, or for which detection SN is easier, i.e face-on galaxies. After studying the properties of 1000 galaxies that have produced SNe since 1885\footnote{see \url{http://www.cbat.eps.harvard.edu/lists/Supernovae.html}} we conclude that the most prolific galaxies include:
\begin{enumerate}
\item Nearby galaxies ($\le$ 40 Mpc) with large angular radius ($\ge$ 4 arcmin)
\item Face-on galaxies
\item Starburst or HII galaxies that have a enhanced stellar formation
\item Spiral galaxies or merging galaxies
\end{enumerate}

According to the above criteria, we would expect to detect 7-10 SNe per year from our catalog of 112 galaxies. However, taking into account the variable weather conditions, the moon phases and expected technical difficulties, we should observe 2-3 SN during their early phase with good observational conditions. In order to optimise the number of observed galaxies per night we specify an exposure time for each galaxy to detect a SN to a limiting absolute magnitude of  $R_{abs}\sim$ -13. This limiting magnitude is deep enough to detect faint sources like young SNe without consuming too much observation time.

\begin{figure*}[t]
\begin{minipage}{0.9\linewidth}
\includegraphics[scale=0.38]{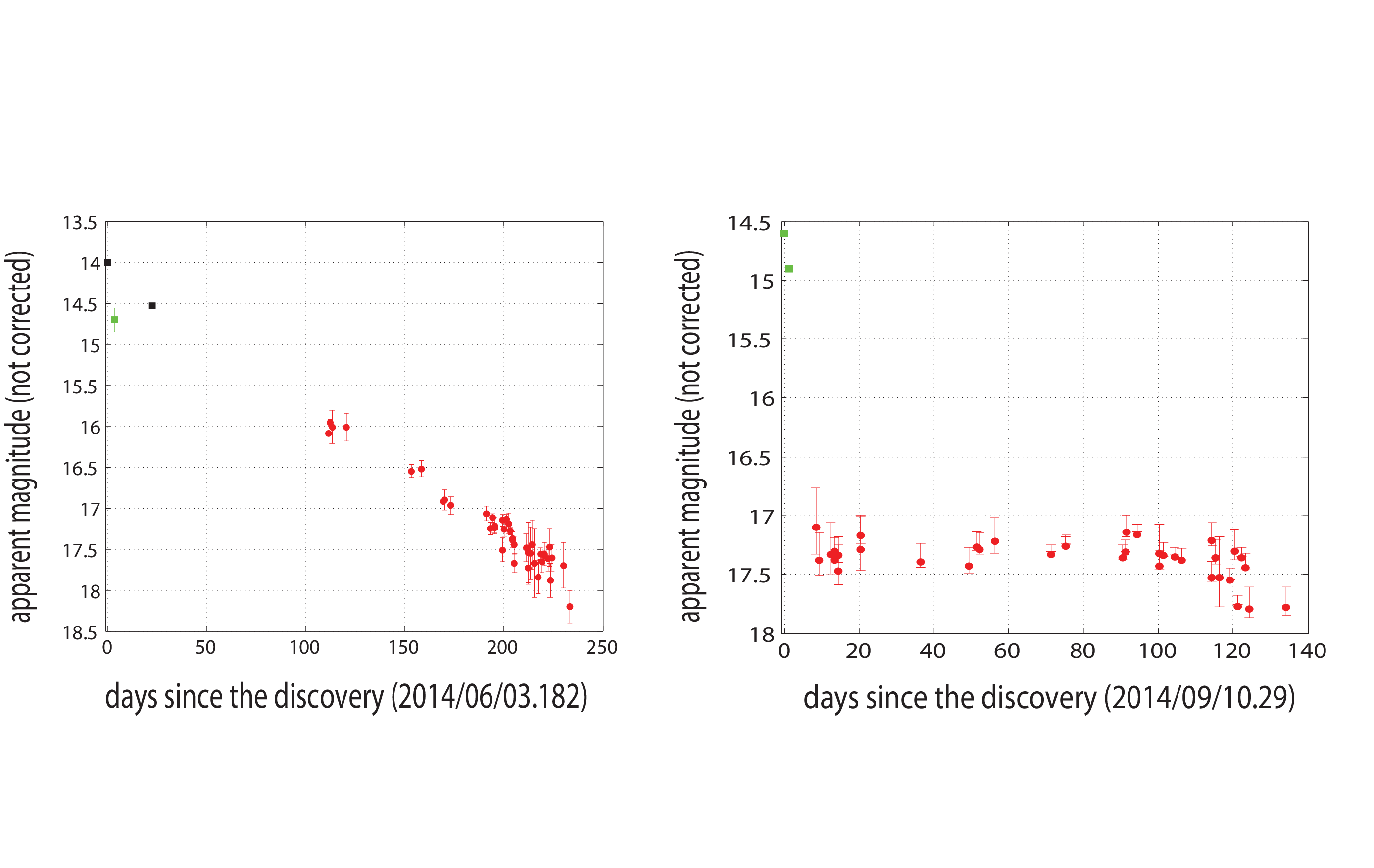}
\end{minipage}
\caption{Light curves of SN2014df (left) and SN2014-ha (right) with R-band (red), V-band (green), and unfiltered observations (black). Photometry is not corrected for Galactic dust extinction. The Zadko measurements are shown as filled circles, observations from other telescopes are represented by filled squares.
}
\label{fig:SN2014df}
\end{figure*}

The program was implemented in June 2014 and in 2014-2015, 8 SNe were discovered in galaxies belonging to our catalog : SN2014A (NGC5054), SN2014B (NGC4939), SN2014C (NGC7331), SN2014J (NGC3034), SN2014L (NGC4254), SN2014df (NGC1448), ASAS-SN2014ha (NGC1566) and SN2014dt (NGC4303). Among them 6 were classified as CCSNe and 2 as Type Ia SNe. Unfortunately, the first 5 SN were discovered 4 months before the program commenced. Due to observational constraints, the Zadko Telescope detected the 3 other supernovae (SN2014df, ASAS-SN2014ha and SN2014dt) days after their discovery (see Fig. \ref{fig:SN2014df} for light curves of SN2014df and SN2014ha).
Example images can be found at \url{http://www.rochesterastronomy.org/sn2014}, \url{http://www.astronomerstelegram.org/?read=6460} and \url{http://www.rochesterastronomy.org/sn2014/}.


Since the beginning of the SN program, the number of SN discovered per galaxy is in good agreement with the expected rate. So far, the project has not detected an early SN, which we attribute to a poor duty cycle from weather conditions, technical upgrades and observational constraints. The project will continue to search for young and nearby SNe with a future update of the targeted galaxy catalog.

\section{Optical follow-up of gravitational wave triggers}
\label{sec:gw}
The first detections of gravitational waves (GWs) from coalescing black hole binary systems: GW150914 \citep{PRL} and GW151226 \citep{2016PhRvL_GW151226}), in late 2015 opened up a new observational window to the Universe. As GWs propagate virtually freely through space, they can provide a pathway to dynamic astrophysical processes presently inaccessible by electromagnetic (EM) detection. Among the strongest emitters in the GW spectrum are cataclysmic sources, such as coalescing systems of compact binaries and the collapse of massive stars. Future detections may have associated emissions in the EM spectrum allowing the opportunity for coordinated multi-wavelength observations.

The GW detector network is being led by Advanced LIGO \citep[aLIGO;][]{Abadie2010CQGra} and will be followed by Advanced Virgo \citep[AdV][]{2015CQGra..32b4001A} in 2017. When these advanced detectors reach design sensitivity (around 2019), they will have 10 times greater sensitivity than the first generation instruments (2003-2009) yielding a factor 1000 improvement in observed volume. These facilities can detect coalescing systems of neutron stars to distances approaching $\sim 400$ Mpc, neutron star--black hole systems to $\sim 1$ Gpc and stellar mass black hole coalescences like GW150914 out to cosmological distances greater that $z \sim 1$.

The coordinated follow-up of GW events in the EM spectra and beyond will increase the confidence of detections, maximise the scientific output and allow more sensitive searches into GW data, extending the sensitivity horizon. As the detector network sensitivity improves incrementally in stages towards design sensitivity, we expect to probe other source populations, such as binary neutron star and neutron star - black hole mergers. Such joint detections will initiate a new era of multi-messenger astronomy.
\subsection{The aLIGO/AdV follow-up program}
For the first aLIGO observation run (O1 which ran from Sept 2015 to Jan 2016), Zadko was one of a number of EM facilities selected for follow-up GW sources. The LIGO/Virgo collaboration had over 63 worldwide EM partners with signed MoUs for conducting follow-ups of GW triggers at that time; around 70 MoUs will be in place for the second observation run in late 2016 (O2). Of the partners, 73\% are in the UV/Optical/IR, 11\% in the radio and the remainder are in high energy X-ray and Gamma-ray. Australian involvement includes MWA \citep{Tingay2013PASA}; ASKAP (VAST) \citep{Murphy2013PASAa}; AAT \citep{Tinney04iris2:a}; SkyMapper \citep{Keller2007PASA},GOTO\footnote{\url{http://goto-observatory.org/}} and the ``Deeper Wider Faster" project which targets simultaneous, fast cadenced observations through shadowing of optical and radio fields; interesting candidates are followed up using a number of partner EM facilities, including Zadko, covering low to high energy \citep{Howell2015PASA}. In the high energy gamma-ray there will be Australian involvement through the Cherenkov Telescope Array \citep[CTA;][]{Acharya_2013APh}, The High Energy Stereoscopic System \citep[H.E.S.S;][]{Lennarz2013}, as well as coordinated neutrino observations using IceCube \citep{IC2006APh}.

The main latency bottleneck during O1 was the automatic and manual verifications which included checks on data quality and conditions. Therefore, to allow alerts to be sent out significantly faster, automation is a particular aim for future science runs in the advanced GW detector era. The statistical significance of a GW trigger is given through its false alarm rate (FAR), which is the rate that false positives appear above a given signal-to-noise ratio threshold. Typical FARs for triggers sent out during the O1 were of order 1/month.

To rapidly communicate the information on GW events the VOEvent\footnote{\url{http://www.ivoa.net/documents/VOEvent/}} standard was adopted \citep{Williams2012SPIE}. The content of a VOEvent alert sent out by aLIGO for a compact binary coalescence event included estimates of the FAR (in Hz) and the sky position of the source provided by way of a probability sky map (which can be multi-modal and non-Gaussian). Future runs could include additional information such as estimates of the luminosity distance of the source and an indication of the component masses of the system.

\subsection{The first EM follow-up of a GW transient}
The first confirmed gravitational-wave transient, GW150914 (a binary black hole merger) was discovered by the aLIGO detectors on
2015 September 14 \citep{PRL}. Preliminary estimates of the time, significance, and sky location
of the event candidate were shared with 63 teams of observers covering radio, optical, near-infrared, X-ray,
and gamma-ray wavelengths with ground- and space-based facilities.  The low-latency analysis
of the gravitational wave data produced web accessible sky position maps. Of the 63 teams, 25 follow-ups were reported via a private Gamma-ray Coordinates Network Circulars. The follow-up observations (by 25 teams) for this first campaign is outlined in \citet{2016ApJGW150914FuP}; this included a disputed claim of a possible short gamma ray burst coincident (within 60 ms of GW150914) with the GW event \citep{2016ApJFermi}. Despite over 60 participating facilities with MoUs for EM follow-up, around a third managed to acquire data for this first event. This can be attributed to many factors, including weather (optical), instrumental problems and rapid follow-up not possible because of longitude separation. For the first follow-up campaign, the only Australian facilities or collaborations to obtain follow-up data were SkyMapper and TZAC (TAROT - Chile), of which Zadko (the Australian node) was not operational at the time of the trigger.

\subsection{Gravitational wave follow-up challenges and strategies}
Looking ahead to the second aLIGO/Virgo observation run (02; late 2016), the most likely GW sources to yield optical counterparts are the mergers of coalescing binary systems of neutron star. As discussed in section \ref{section_GRBs}, these events have been strongly suggested as the progenitors of short hard $\gamma$-ray bursts \citep{Tanvir2013Natur,Berger2013ApJ}. The evidence stems from the fact that short-hard GRBs are often observed in older stellar populations with offsets of tens of kpc from their galactic centers; this is consistent with binary compact object formation channels. Additionally, follow-up observations in the infrared of GRB 130603B provided evidence of a `kilonova'; a faint transient predicted to form after the merger of two neutron stars, and powered by the radioactive decay of the ejected neutron rich matter \citep{Li1998ApJ,Rosswog2005ApJ,Metzger2010MNRAS}.

As highlighted during O1, a particular challenge for optical follow-ups of GW sources during O2 and beyond will be error boxes of 100s deg$^{2}$; compare this with the typical error boxes for GRB follow-ups of order arc mins. As mentioned earlier, these error regions are provided to optical facilities as probability sky maps. The morphology of the skymaps is dependent on the location of the source in the sky relative to the antenna pattern function of the GW detector network. The shapes of the probability map can take the form of a single elongated arc which could cover several hundred square degrees, or two or more degenerate arcs which could include small isolated regions. When Virgo increases the network to 3-detectors (in 2017) localisations should improve; one can expect confidence regions of 10s of degrees in less than 17\% of events \citep{Singer2014ApJ}.

To make the most of opportunities, instrument dependent strategies must be employed to cover the most probable regions of the error region in sufficient time to capture a fading EM source. These strategies will have to be adapted depending on the particular morphology and for partially accessible error regions, the component available to the southern hemisphere. The most likely EM counterparts, short hard GRBs have fainter afterglows than the longer duration bursts. However, for on-axis sources within the aLIGO/AdV horizon one would expect these events to be bright enough for identification; however event rates and beaming considerations suggest that in most cases the bursts would be off-axis requiring deep searches.

Different follow-up techniques to screen candidate counterparts are still in the testing stage; these include simple tiling strategies employing galaxy catalogues to more sophisticated clustering algorithms based on machine-learning. Source confusion will also be an important consideration. Sources such as Solar System objects can be excluded by having a cadence of order 10s of mins between images of the same field and supernovae can be excluded by their relatively long durations.

Initial follow-ups by Zadko will employ tiling the highest probability sky regions and secondly target only the brightest galaxies utilizing the current automated catalogue matching pipeline that creates a list of unknown candidates for an alert image\footnote{From the GW Candidate event Database, GraceDB; \url{https://gracedb.ligo.org/}}. Image subtraction can also be used for galaxies that we have reference images i.e. from our current supernovae search.

\subsection{Coordinated GW follow-up imaging with Zadko}

With the intense interest in GW follow-up, a number of facilities are being constructed as dedicated instruments with fields of view (FoVs) large enough to cover the GW error regions in reasonable times. One such facility that could be operational during the later stages of aLIGO/AdV is RAMSES; if successful a similar instrument could operate alongside Zadko. The RAMSES prototype will consist of a set of 16$\times$40cm telescopes equipped with rapid cameras capable of readout times from 100ms to several minutes; it will be able to reach magnitudes of 15 in 1s and mag 20 within around 10 mins. Each telescope has a 2.5 deg$^{2}$ FoV, resulting in a combined 100 deg$^{2}$ FoV; this in combination with a slew time of around 5s would enable instantaneous coverage of a large proportion of a typical aLIGO/AdV error region. RAMSES has received funding and will be based in National Aures Observatory in Algeria\footnote{\url{http://www.craag.dz/}} with a prototype built in France.

Comparable instruments employed in GW follow-ups include BlackGEM \citep[40deg$^{2}$ FoV; mag 22 in 5m;][]{Bellm_2014} and the
Zwicky Transient Facility \citep[ZTF; 47deg$^{2}$ FoV; mag 20.5-21 in 30s;][]{Ghosh2015}; other than being more sensitive, these telescopes operate as stand-alone facilities, therefore a telescope like RAMSES could have a niche role alongside a more sensitive instrument. The wide FoV and rapid slew of RAMSES allows for rapid scanning of the GW error region for bright transients which can then be targeted for instantaneous deep follow-up by the Zadko telescope.

We plan to build and install RAPIDO (very rapid slewing mount co-located with Zadko) using two wide field telescopes combined with a CCD imager that will achieve $m < 17$ with FoV (8-10)deg$^{2}$. The single telescope system, comparable to RAMSES but using off the shelf optics, is expandable, and can accommodate either 0.4m or 2 x 0.2m telescopes on each mount. The system will provide low latency wide field searches for transients, and will automatically send alerts to Zadko for deep follow-up ($m < 21$) of any transient candidate. We note that when Virgo joins aLIGO, for the brightest GW sources the localisation error regions will reduce to some tens of degrees, providing opportunities for low-latency imaging of the entire field by RAPIDO and deep follow-up by Zadko.

\section{Fast Radio Burst optical follow-up}

The discovery of fast radio bursts (FRBs) revealed the existence of an exciting new class of astronomical object \citep{lor07}. FRBs are a transient phenomenon of Jy level flux found in GHz bands with a very short duration of a few ms. FRBs are very difficult to localise, as they are mostly detected by single radio dishes (more than a dozen from Parkes and one from Arecibo) with fields of view (FoV) of many arc-minutes. As of Feb 2016, no repeating events or counterparts in other wavelengths had been publicly announced \citet[e.g. for FRB140514 as reported in][]{pet15}.  The first localisation, made public in Feb 2016 \citep[FRB 150418;][]{k16}, was achieved by the discovery of a fading radio transient, which enabled deep optical imaging to identify the host galaxy at $z\approx0.5$. Identification of an afterglow and host was later called into question as the source was identified as an AGN \citep{2016ApJ...821L..22W}. Furthermore, observations of FRB 110523 using archival data from the Green Bank Telescope \citep{masui15} favour models involving young stellar populations such as magnetars, over models involving the mergers of neutron stars, which are more likely to be located in low-density regions of the host galaxy. These contradictions can be resolved by obtaining multiple coincident multi-wavelength EM (especially optical) counterparts, and to search for possible GW signatures.

\subsection{ Zadko Telescope automated follow-up of Parkes}\label{zfrb}
\begin{figure}
\centering
\includegraphics[scale = 0.45, angle = -90 ]{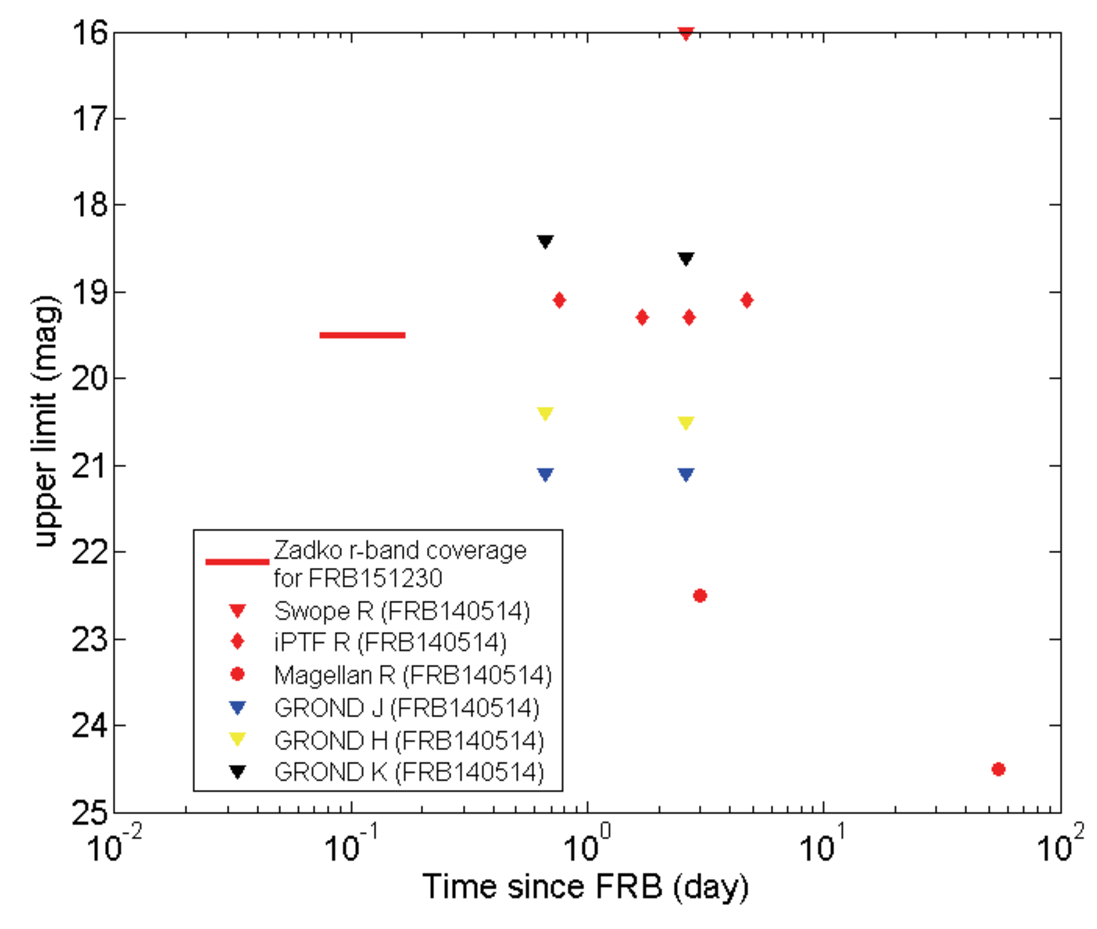}
\caption{ Zadko on source delay time (horizontal left line) for FRB follow-up, compared to all other optical follow-ups (symbols). The low latency FRB response was obtained during a Parkes {\it shadowing} experiment in Dec 2015, which set the record for the shortest time delay to image the FRB alert. The Zadko limiting magnitude of 19.5 (60 s exposure in r) is improved by about 0.8 mag in the stacked image. A complete analysis of this data will be presented in a subsequent work.} \label{frb}
\end{figure}

The Zadko Telescope and in the future RAPIDO telescopes, based on their reaction times and sensitivities for GRB imaging, have the potential to fill a niche role in extremely rapid follow up of radio transients events (see Figure \ref{frb}). Previously, the most rapid follow up of FRB140514 was 7 hours later, with ATCA radio telescope. The most rapid optical follow up was performed by GROND (Gamma-Ray Burst Optical/Near-Infrared Detector) but this was 16 hours post alert. SWOPE \footnote{\url{http://obs.carnegiescience.edu/swope}} observations, also more than 16 hours after the event, provided a limiting magnitude of only $R=16$. None of these follow-ups could offer a candidate for the host galaxy.

As of April 2015, the Zadko Telescope joined a large collaboration which is searching for FRBs in realtime. The project is part of the alert network for the SUPERB\footnote{https://sites.google.com/site/publicsuperb/fast-radio-bursts} and UMOST\footnote{http://astronomy.swin.edu.au/research/utmost} projects. As part of this effort the Zadko control software has been updated to respond robotically to FRB alerts. Results from these observations have the potential to probe a region of parameter space not covered.

In December 2015, the team commenced a pilot project for several weeks that enabled the Zadko Telescope to shadow (follow) the same sky location as Parkes. Because the Parke's beam is about 4 times the area of Zadko's FoV, we employed a tiling strategy. This first Zadko pilot experiment broke the record for the fastest response (about 40 min post burst), and earliest optical data following an FRB alert (see Fig \ref{frb}). The first imaging of the source occurred before the human vetted alert was sent out from Parke's (analysis of this low latency data is on-going and will be reported in a subsequent work).

Based on event rate estimates from the 2015 observing sessions, 10 FRBs should be discovered in Parkes data in 2017. This provides an opportunity to search for a very early optical counterpart, to identify a possible short GRB afterglow, localise a host galaxy to confirm the distance, and to study the host environment.

\vspace{-0.02cm}

\section{Education, outreach and training}
\subsection{Science education enrichment}
The Zadko Telescope has been used for several novel science education programs. Local PhD student projects include the evaluation of authentic learning by participation in research projects \citep{2012cosp...39..735H,2011AdSpR..47.1922C}. These projects include local school students identifying and searching for near Earth asteroids using the Zadko Telescope. This was extended as part of the Yachad visiting scholar award (led by Coward in November 2012) who initiated a pilot scheme that provided local school students in Israel access to the UWA Zadko Telescope for participating in the above project. In addition there has been numerous undergraduate student training programs, both hosted locally and remotely from France, that have used the Zadko Telescope for their projects.

\subsection{Space science training: space debris}

The potential exponential growth of artificial debris in Earth orbit together
with its consequent hazard to satellite operations \citep{kenbrock07}.
has been known for some time (Kessler and Cours-Palais, 1978).  Most observations
and studies have concentrated on the low Earth orbital regime (LEO).  There are several military and civilian observing patrol networks that contribute to the maintenance of catalogs of the larger debris sizes ($>10$ cm).  Occasional scientific campaigns are undertaken to explore smaller sized debris, as it has been realised that the growth
of even micron sized debris has the potential to interfere with astronomical
observations in the medium future \citep{bigken}.

Collisional velocities in geosynchronous Earth orbit (GEO) are lower than
those in LEO, and this has led, along with the greater difficulty of
observation, to a lesser interest in GEO debris.  However, the discovery a
few years ago of a previously unknown GEO debris population \citep{2004AdSpR..34..901S,Schild07} has led to the realisation that the hazard in this region has been considerably underestimated.  This high altitude region is totally beyond the range
of radars used for LEO patrol, and metre-class optical instruments are required
for useful analyses.

The Zadko telescope has previously been involved in
imaging Australian space debris \citep{myrt}
and we currently have an open PhD project to characterise the GEO debris
population in this longitudinal region of the geosynchronous orbit.
(One of us has been involved in a study with Kitt Peak National Observatory
that has shown that the large space object population at GEO
shows a significant longitudinal variation.)

The substantial public and schools interest in artificial space debris
(space debris is a component of the WA school science curriculum) offers a
great opportunity for public outreach in this field as well as for training
of teachers and students in the acquisition of hands on skills in planning
and implementing telescope scheduling and debris database access and analysis.
These activities are expected to be coordinated with a co-located node of
the Falcon Telescope Network dedicated space surveillance facility
\citep{chun}.

%
%

\section{Summary and emerging priorities}
The Zadko Telescope has been operating as a fully robotic optical facility prioritised for automated transient source photometry from March 2010. From 2012 to 214, the facility underwent a major upgrade with the installation and commissioning of an automated observatory building. Despite lengthy disruptions to the operation during this period, the Zadko Telescope emerged as the premier Australian facility for automated triggered follow-up of gamma bursts (40 follow-ups) and neutrino localisations (58 follow-ups). In the context of GRBs, the Zadko Telescope has placed constraints on a possible GRB origin for 2 neutrino events that were rapidly followed up (ANT140301A and ANT141220A).

Solar System science has been an important component of the overall Zadko science program. From 2014 to 2015, the telescope contributed to accurate period determinations for 5 asteroids by obtaining critical light curve measurements. The accuracy of relative photometry by Zadko exceeded the project requirements i.e. dispersion $<$0.01 magnitudes for targets of relatively high brightness (V$\sim$11-13), despite the poor seeing at the site.

Although GRB and neutrino follow-up will continue to be a priority for the transient source program, there are at least two emerging fields that will become a priority for the facility. Rapid follow-up of FRB triggers from Parkes has the potential for breakthrough science, given the debate on the origin of FRBs. This first Zadko pilot experiment broke the record for the fastest response and earliest photometric limits from an FRB alert (Figure \ref{frb}). The first imaging of the source occurred before the human vetted alert was sent from Parkes (analysis of this low latency data will be presented in a later study).

Optical follow-ups of GW alerts from aLIGO have the potential to unveil the physics of neutron star mergers. Given the poor localisation of aLIGO (which will improve when joined by Virgo), we are planning an upgrade to the Zadko facility. It will consist of co-locating a very rapid response wide FoV two telescope system that will be used to scan much of the GW error region. Transient candidates identified by the wide FoV system will be automatically, and rapidly followed up with deep imaging by the Zadko Telescope. When Virgo joins aLIGO are at design sensitivity, the brightest GW sources will be localised to some tens of degrees, allowing low-latency imaging of the entire field by the RAPIDO Zadko facility.

\begin{acknowledgements}
The Zadko Telescope was made possible by a philanthropic donation by James Zadko, Director of Clair Energy, to the University of Western Australia (UWA). In addition, we thank the West Australian Government (Department of Environment and Conservation), Perth Observatory, Australian Research Council grant (LE12010051) and the University of Western Australia for financial support for the upgraded Zadko Observatory. D.M. Coward was supported by the Australian Research Council Future Fellowship (2011-2015 FT100100345) and acknowledges advice from S. Driver and G. Meurer at the International Center for Radio Astronomy Research at UWA. B. Gendre acknowledges financial support from the NASA Award NNX13AD28A and the NASA Award NNX15AP95A. EJH acknowledges support from a UWA Research Fellowship. We also thank Andrew Burrell (School of Physics UWA), and France Romain Laugier for technical support and finding solutions to the many technical problems.
\end{acknowledgements}


\end{document}